# From Access to Realized Affordances: University Students' Generative AI Engagement across Linguistic and Sociotechnical Contexts

1. Ming Li, Institute for Transdisciplinary Graduate Degree Programs, The University of Osaka
2. Qin Xie, Comparative and International Development Education, University of Minnesota,
3. Ariunaa Enkhtur, Institute for Global Initiatives, The University of Osaka
4. Lilan Chen, Center for Higher Education Studies, Waseda University
5. Fei Cheng, Graduate School of Informatics, Kyoto University; National Institute of Informatics

## Abstract

Generative artificial intelligence (GenAI) is becoming increasingly embedded in university students' academic work, yet student engagement is often examined through adoption, frequency of use, or general perceptions, with less attention to how it is shaped by linguistic and sociotechnical conditions. This comparative qualitative study examines how university students access, incorporate, and evaluate GenAI across three contrasting higher education settings in China, Japan, and Mongolia. Semi-structured interviews were conducted with 42 undergraduates from one university in each setting, and the data were analyzed through qualitative interpretation supported by co-occurrence network and correspondence analyses using KH Coder. Across the three samples, GenAI was incorporated into academic practices spanning exploration, production, and refinement. However, access pathways differed across the three contexts. Participants in China navigated between domestic and global GenAI systems, those in Japan predominantly used globally available systems through Japanese, and those in Mongolia more frequently shifted toward English or other languages when Mongolian-language outputs were perceived as less satisfactory. Students also continued to use GenAI while evaluating its accuracy, privacy implications, potential for dependence, and effects on critical thinking. These findings suggest that technological access alone does not ensure equivalent educational usability across linguistic and sociotechnical contexts. Building on the empirical findings and the analytical perspectives adopted in the study, an empirically informed integrative framework is developed in which student GenAI engagement is understood as situated, mediated, enacted, and evaluated. The framework conceptualizes realized affordances as educational possibilities that become practically usable through the interplay of sociotechnical conditions, linguistic resources, academic practices, and evaluative judgments.

Keywords: generative AI; higher education; student engagement; realized affordances; linguistic mediation; sociotechnical context

## 1. Introduction

GenAI has become increasingly embedded in university students' academic work, supporting activities such as information seeking, idea development, explanation, translation, writing, and revision (Chan & Hu, 2023; Schei et al., 2024; Stöhr et al., 2024). Yet whether students use GenAI, or how frequently they use it, reveals only part of how these technologies become involved in learning. Research has shown substantial variation in the purposes for which students rely on GenAI and in the roles these tools assume in academic tasks, ranging from information retrieval and feedback to planning and assignment production (Stojanov et al., 2024; Schei et al., 2024). Equally important, therefore, is understanding how GenAI is incorporated into academic work, including where it enters students' learning processes, what purposes it serves, and how its outputs are taken up, evaluated, or revised. The key question is not simply whether students use GenAI, but how its technological possibilities are enacted within academic practice.

Yet the enactment of GenAI in academic practice does not occur under uniform conditions. Students encounter GenAI within different sociotechnical environments shaped by platform availability, local AI ecosystems, and linguistic resources (Li et al., 2026). In non-English-speaking higher education, language is particularly consequential. It is not only a target of GenAI use, as in translation or writing support, but also a medium through which students interact with AI systems, formulate prompts, interpret responses, and judge their usefulness and reliability (Ou et al., 2024; Wang, 2024). Platform access and language may also operate together: research in China, for example, shows that accessibility, language proficiency, disciplinary needs, and perceived platform capabilities influence students' choices between domestic and global GenAI systems (Xie et al., 2025). These findings point to an important distinction between formal access to GenAI and the educational affordances that students are actually able to realize through its use.

China, Japan, and Mongolia provide analytically useful contrasts for examining how these conditions shape students' engagement with GenAI. Although all three are non-English-speaking higher education contexts, they differ in the combinations of platform access, local AI ecosystems, and linguistic resources that students encounter. In China, a rapidly expanding domestic GenAI ecosystem coexists with differentiated access to some globally dominant platforms, requiring students to navigate between domestic and global systems (Xie et al., 2025). Japan presents a different configuration, with relatively direct access to major global GenAI systems alongside extensive opportunities for interaction in Japanese (Mok, 2024; Wang, 2024). Mongolia, by contrast, combines access to global GenAI platforms with a comparatively more limited Mongolian-language digital resource base, making interaction language potentially more consequential (Munkhsaikhan & Dalkhjav, 2025). These contexts are therefore not treated as homogeneous national populations or fixed cultural categories, but as contrasting sociotechnical and linguistic configurations through which GenAI may be accessed and used differently.

Despite the rapid growth of research on students' use of GenAI in higher education, existing scholarship remains fragmented across different aspects of engagement. Studies have often focused on adoption, frequency of use, general perceptions, or particular applications of GenAI, while language-related practices, academic uses, and concerns are frequently examined as separate issues (Chan & Hu, 2023; Schei et al., 2024; Stojanov et al., 2024; Ou et al., 2024; Wang, 2024). What remains underdeveloped is an integrative account of how platform conditions, linguistic resources, academic practices, and students' evaluative judgments operate together in situated engagement with GenAI. Such an account is needed to connect the conditions under which GenAI is accessed with how it is mediated through language, enacted in academic work, and evaluated in relation to its benefits and risks. This perspective also makes it possible to examine how similar forms of technological access can be accompanied by different realized educational affordances across contexts.

Against this background, this study examines how university students engage with GenAI across contrasting linguistic and sociotechnical contexts. Drawing on semi-structured interviews with 42 students from one university in each of China, Japan, and Mongolia, supported by computer-assisted text analysis, the study investigates students' choice of GenAI tools and interaction languages, how GenAI is incorporated into their academic practices, and the concerns they associate with its use. It further develops an integrative four-dimensional framework that conceptualizes GenAI engagement as situated within sociotechnical conditions, mediated through linguistic resources, enacted through academic practices, and evaluated through students' judgments of benefits and risks. By shifting attention from technological access alone to the educational possibilities students are actually able to realize, the study highlights the importance of understanding GenAI engagement as a context-dependent and linguistically mediated educational practice. This distinction is particularly important for higher education institutions seeking to integrate GenAI in ways that are not only technologically accessible but also educationally meaningful across diverse linguistic contexts.

## 2. Literature Review

### 2.1 Student GenAI Engagement as Academic Practice

GenAI has become increasingly integrated into university students' academic activities. Research documents a broad range of uses, including information seeking, explanation, brainstorming, summarization, translation, academic writing, feedback, programming, and problem solving (Chan & Hu, 2023; Schei et al., 2024; Qian, 2025). Recent reviews further suggest that GenAI is assuming increasingly differentiated roles within learning rather than functioning as a single-purpose tool. For example, An et al. (2026), reviewing 39 empirical studies, identified seven broad ways in which higher education students used GenAI, including as a translator, refiner, navigator, evaluator, dialoguer, creator, and supporter of self-regulated learning. Such evidence indicates that GenAI is becoming embedded in academic workflows and that student engagement cannot be adequately represented by a simple distinction between use and non-use.

Recent research has consequently begun to move beyond adoption and frequency toward the different ways students engage with GenAI. Stojanov et al. (2024), for instance, identified distinct reliance profiles ranging from knowledge seeking and proactive learning to assignment delegation. Yang et al. (2024) approached the issue through student agency and distinguished receptive, resistive, resourceful, and reflective forms of GenAI-supported learning. Hou et al. (2025), focusing on problem solving, similarly identified reflective, cautious, thoughtless, and collaborative forms of reliance. Qualitative studies have also shown that students may incorporate GenAI at different stages of an academic task and combine it with other digital tools. Zhao et al. (2024), for example, found that students used ChatGPT alongside other technologies in individualized ways within the iterative process of academic writing. Together, these studies demonstrate that GenAI engagement is heterogeneous and processual, involving different relationships among learners, technologies, and academic tasks.

Student engagement with GenAI also involves continuing evaluation of its benefits, limitations, and consequences. Existing research has documented concerns about inaccurate or fabricated information, privacy and data security, academic integrity, authorship, overdependence, and possible erosion of independent or critical thinking (Gruenhagen et al., 2024; Li et al., 2024; Li et al., 2025; Qian, 2025). These concerns are closely connected to actual patterns of use. The same capabilities that enhance efficiency may also encourage the outsourcing of cognitive work; similarly, generative support may facilitate writing and problem solving while raising questions about authorship, responsibility, and the extent to which students remain intellectually engaged. Recent research therefore increasingly examines how students engage with GenAI through patterns of reliance, agency, collaboration, and judgment rather than through simple adoption alone.

This growing literature establishes an important basis for understanding GenAI as part of academic practice. At the same time, much of the process-oriented evidence remains concentrated in particular tasks, such as academic writing, language learning, and structured problem-solving activities, or within individual institutional and national settings. Less is known about how students incorporate GenAI across a broader range of everyday academic practices, potentially using multiple systems, and how these patterns take shape under contrasting technological and linguistic conditions.

### 2.2 GenAI Engagement across Sociotechnical and Linguistic Contexts

Student GenAI practices are shaped by the technological environments in which they occur. Access is more complex than a binary distinction between having or not having a GenAI tool. Students may encounter different combinations of globally available systems, locally developed alternatives, subscription requirements, regulatory conditions, and platform-specific capabilities. These conditions can influence both which systems are accessible and which tools students select for particular purposes. Research in China illustrates this complexity. Xie et al. (2025) found that university students navigated between domestic and global GenAI systems and that their choices were shaped by accessibility, language proficiency, disciplinary needs, and perceived platform capabilities. Tool choice therefore forms part of a broader platform ecology through which students encounter and incorporate GenAI into academic work.

Language introduces a further layer of complexity. A substantial body of research has examined GenAI as a resource for translation, academic writing, linguistic revision, and foreign-language learning (Ou et al., 2024; Wang, 2024). More recent scholarship has also begun to examine multilingual forms of human–AI interaction. Translanguaging perspectives highlight how multilingual learners may draw on different elements of their linguistic repertoires when interacting with GenAI rather than remaining within a single named language (Ahuja et al., 2023). At the same time, multilingual functionality should not automatically be equated with linguistic inclusion because generative systems may reproduce linguistic hierarchies and underrepresent less dominant languages and forms of expression (Jeon et al., 2025).

This distinction points to two analytically different roles of language in GenAI-supported education. The first concerns GenAI for language. Students use AI for translation, language correction, foreign-language learning, and academic writing. The second concerns language in GenAI engagement. In this case, language shapes how students formulate prompts and interpret responses. It also influences how they assess output quality and decide how to continue interacting with a system. This issue is important because multilingual GenAI systems do not necessarily perform equivalently across languages. Gupta et al. (2025), evaluating major

LLMs on educational tasks across nine languages, found that performance varied partly in relation to the amount of linguistic representation in training data, with lower-resource languages tending to perform less well on some tasks. Broader language-model research similarly shows that performance is influenced by the quantity and composition of linguistic resources available across languages (Chang et al., 2024).

Such differences in model performance may become educationally consequential when students adapt their interaction strategies. They may switch languages, reformulate prompts, combine GenAI with translation technologies, or move between systems in response to perceived output quality. Linguistic repertoire can therefore become a practical resource for engaging with GenAI. English proficiency, for example, may function not only as an object of language learning but also as an additional resource for interacting with globally developed AI systems.

China, Japan, and Mongolia provide useful contrasting settings for examining the intersection between platform ecology and linguistic resources. In China, a substantial domestic GenAI ecosystem coexists with differentiated access to globally dominant systems, creating conditions in which tool selection can itself form part of students' AI practices (Xie et al., 2025). In Japan, major global GenAI systems are relatively accessible, and existing research shows that students can use them for academic and linguistic support through both Japanese and English (Mok, 2024; Wang, 2024). Mongolia represents a less extensively studied setting.

Emerging research indicates that ChatGPT is entering university students' learning activities (Khashkhuu, 2025; Munkhsaikhan & Dalkhjav, 2025), while the comparatively smaller digital resource base for Mongolian makes the linguistic dimension particularly relevant.

These differences should not be interpreted as fixed national or cultural characteristics. Rather, the three settings represent contrasting configurations of platform availability, local AI ecosystems, and linguistic resources. Examining these configurations makes it possible to consider how sociotechnical and linguistic conditions are reflected in students' actual GenAI engagement.

### 2.3 Research Gaps and Research Questions

The literature reviewed above shows that research on student GenAI engagement has moved beyond early questions of adoption and general perceptions. Process-oriented studies increasingly demonstrate variation in how students rely on GenAI, exercise agency, collaborate with AI systems, and incorporate them into different stages of learning. Nevertheless, three related gaps remain.

First, much of the existing process-oriented research focuses on particular tasks or pedagogical settings, especially academic writing, language learning, and structured problem solving. These studies provide valuable insights into specific forms of interaction but offer a more limited account of how GenAI is incorporated across students' broader everyday academic work. Students may use GenAI for information seeking, comprehension, idea development, writing, translation, summarization, revision, and multimodal activities, sometimes moving between several functions or systems within the same academic workflow. Comparative evidence on such broader patterns across contrasting higher education settings remains limited.

Second, although the linguistic dimension of GenAI engagement has received growing attention, less is known about how students respond in practice to differences in linguistic usability across AI systems. Research on language learning, multilingual writing, and translanguaging has shown how students mobilize linguistic resources in AI-supported learning, while computational studies have demonstrated variation in GenAI performance across languages. What remains less well understood is how students negotiate these conditions through language switching, prompt reformulation, translation, or tool selection as part of their wider academic use of GenAI.

Third, platform conditions, linguistic resources, academic practices, and students' evaluative judgments are often examined separately. Yet these dimensions are interconnected in students' actual experiences. Platform conditions shape which systems are available, linguistic resources influence how students interact with them, academic tasks affect how AI capabilities are put into practice, and experiences with useful or problematic outputs inform subsequent judgments. More integrated empirical accounts are therefore needed to examine how these dimensions come together across contrasting higher education contexts.

China, Japan, and Mongolia provide analytically contrasting settings for examining these relationships. The study does not treat the three countries as homogeneous or nationally representative populations. Instead, they are used as contrasting higher education contexts characterized by different configurations of platform ecology and linguistic resources.

Accordingly, the study addresses the following research questions:

RQ1. How do university students in China, Japan, and Mongolia access and interact with GenAI, particularly in terms of tool choice and interaction language?

RQ2. How is GenAI incorporated into students' academic practices, and what major patterns and cross-context variations emerge?

RQ3. What major concerns regarding GenAI use emerge from students' accounts, and how do these concerns vary across the three study contexts?

## 3. Analytical Perspectives

This study draws on several complementary analytical perspectives to interpret students' engagement with GenAI. These perspectives are not treated as a priori categories into which the interview data are fitted. Rather, they provide conceptual resources for examining different aspects of the empirical patterns identified through the qualitative and computer-assisted analyses. In particular, the analysis is informed by sociotechnical perspectives on situated technology use, sociocultural understandings of linguistic mediation, relational conceptions of technological affordances, and scholarship on critical evaluation within AI literacy. Together, these perspectives help direct attention to the conditions under which students encounter GenAI, the linguistic resources through which interaction occurs, how technological possibilities are incorporated into academic activities, and how students judge the usefulness and limitations of AI assistance.

A sociotechnical perspective provides a basis for examining the conditions under which students encounter and use GenAI. Orlikowski's (2000) practice lens emphasizes that technology use is situated and acquires meaning through its enactment within particular configurations of resources, constraints, and social practices. Building on this perspective, the present study attends to the technological environments in which students make choices about GenAI, including the accessibility of particular platforms and the availability of domestic and global systems. These features are treated as characteristics of the settings in which technology use occurs rather than as outcomes of "country" itself. China, Japan, and Mongolia are therefore approached as contrasting sociotechnical settings rather than as fixed national or cultural categories.

A second perspective concerns linguistic mediation. Sociocultural theory conceptualizes language as an important mediational resource through which interaction, meaning-making, and learning are organized (Lantolf, 2006). The present study extends this perspective to human–GenAI interaction by examining the linguistic resources students draw on when formulating prompts, interpreting responses, switching between languages, and combining GenAI with translation tools. Language is thus considered not only as an object of AI-supported activity, such as translation or writing, but also as part of the interaction through which students make use of GenAI. This perspective is particularly relevant to a comparison across multilingual settings in which students may technically access similar systems while interacting with them through different linguistic repertoires.

The analysis also draws on the concept of technological affordances to examine how technological possibilities acquire significance within academic activity. Antonenko et al. (2017) emphasize the importance of considering relationships among technological affordances, users' needs, and their abilities when examining educational technology use. Applied here, this perspective avoids treating functions such as text generation, summarization, explanation, or translation as having a fixed educational meaning. Their significance depends on how students take them up in relation to particular purposes and tasks. The affordance perspective therefore helps interpret why similar technical capabilities may be incorporated differently into information seeking, idea development, writing, translation, revision, or other forms of academic work.

Finally, students' accounts are interpreted with attention to evaluative judgment**.** AI literacy scholarship emphasizes that meaningful engagement with AI involves not only use but also critical evaluation and ethical awareness (Long & Magerko, 2020; Ng et al., 2021). In this study, AI literacy is not treated as a measured construct or as an attribute assigned to individual participants. Rather, this literature provides an interpretive resource for examining how students assess issues such as reliability, privacy, dependence, authorship, critical thinking, and the appropriateness of AI assistance. Such judgments are relevant to understanding how students decide whether to accept, modify, verify, limit, or avoid particular forms of GenAI use.

These perspectives are used as complementary interpretive lenses rather than as a fixed or sequential model of student GenAI engagement. The empirical categories and cross-context patterns presented in the findings are derived from the interview corpus and the analytical procedures described below. The theoretical perspectives introduced here are subsequently used to interpret those patterns and to examine relationships among technological conditions, linguistic interaction, academic practice, and students' judgments. A broader conceptual synthesis of these relationships is developed in the Discussion on the basis of the empirical findings.

## 4. Methodology

## 4.1 Research Design

This study employed a comparative qualitative design supported by computer-assisted text analysis to examine university students' engagement with GenAI across three contrasting higher education settings in China, Japan, and Mongolia. Semi-structured interviews were used to capture students' experiences of selecting, interacting with, using, and evaluating GenAI, while KH Coder supported the identification of recurring lexical patterns and cross-context associations.

The three settings were treated as contrasting study contexts rather than as statistically representative national populations. Because participants in each country were recruited from a single university, observed differences were interpreted as situated patterns within the present samples rather than as country-level effects.

## 4.2 Participants and Sampling

The study involved 42 undergraduate students enrolled at one leading national university in each country: China (n = 15), Japan (n = 13), and Mongolia (n = 14). Purposive sampling was used to achieve variation in disciplinary background and gender. The sample included 21 students from the humanities and social sciences and 21 from STEM-related disciplines, with 20 male and 22 female participants (Table 1).

Recruitment procedures were adapted to the local context. Chinese participants were recruited through WeChat, Japanese participants through targeted email invitations within university networks, and Mongolian participants through established academic networks. Data were collected from June to August 2024 in China, from November 2024 to January 2025 in Japan, and in August 2024 in Mongolia.

Table 1. Participant characteristics

| | **China** | **Japan** | **Mongolia** |
|---|---|---|---|
| Institutional setting | One leading national university | One leading national university | One leading national university |
| Interview period | Jun-Aug 2024 | Nov 2024-Jan 2025 | Aug 2024 |
| Total | 15 | 13 | 14 |
| Humanities & social sciences | 9 | 6 | 6 |
| STEM | 6 | 7 | 8 |
| Female | 8 | 5 | 9 |
| Male | 7 | 8 | 5 |

### 4.3 Data Collection

Semi-structured interviews explored participants' academic backgrounds, patterns of GenAI use, institutional guidance and support, attitudes toward GenAI, and reflections on its benefits and limitations. Particular attention was given to tool choice, interaction language, academic activities, and concerns regarding GenAI use.

Interviews lasted approximately 60 minutes and were conducted by researchers fluent in the participants' respective languages: Mandarin Chinese, Japanese, or Mongolian. Interview modes differed across study sites. All interviews in China were conducted online via Zoom, all interviews in Mongolia were conducted face-to-face, and in Japan 11 interviews were conducted face-to-face and two via Zoom.

After informed consent was obtained, interviews were audio-recorded and transcribed. Researchers manually checked each transcript against the corresponding audio recording and corrected inaccuracies where necessary. Personally identifying information was removed, and participants were assigned anonymized identifiers using country codes (C, J, and M). The final dataset comprised approximately 43 hours of verified interview material.

Verified transcripts were translated into English by bilingual members of the research team to create a common corpus for cross-context analysis. Where interpretation depended on recurrent or context-specific terms, the English translations were checked against the corresponding source-language passages to reduce potential distortion introduced through translation.

### 4.4 Data Analysis

Data analysis combined computer-assisted text analysis with qualitative interpretation. KH Coder was used to conduct co-occurrence network and correspondence analyses. The computational outputs were treated as exploratory and heuristic aids rather than as stand-alone quantitative findings and were interpreted in conjunction with relevant interview passages.

The English-language corpus was organized by participant and study context, with country retained as an external grouping variable. Co-occurrence network analyses were conducted separately for students' GenAI use and concerns regarding GenAI use. The sentence was used as the unit of analysis, and terms occurring fewer than four times were excluded to reduce the influence of infrequent words.

The GenAI-use corpus contained 3,848 extracted tokens, of which 1,769 were retained after preprocessing. The concerns corpus contained 1,831 extracted tokens, of which 931 were retained. The latter comprised 598 unique word types, of which 524 were retained. The resulting networks were examined for clusters of closely associated terms. These clusters were not treated as automatically generated themes. Instead, the researchers returned to the corresponding interview passages and interpreted the lexical relationships in context to develop thematic groupings of GenAI practices and concerns.

Correspondence analysis was then used to examine cross-context variation. Country (China, Japan, and Mongolia) was specified as the grouping variable, with the sentence as the document unit. A minimum term frequency of four and a minimum document frequency of one were applied. This resulted in 91 eligible words for GenAI use and 47 for concerns. Terms were filtered according to their chi-square association with country, with up to the top 60 terms specified for visualization. The resulting maps were interpreted through the relative positions of words and country categories. These associations were used to identify contextual tendencies within the dataset rather than to infer statistically representative differences between national student populations.

Lexical clusters and cross-context associations were subsequently examined against the original interview passages. Representative quotations were used to contextualize the identified patterns, and thematic interpretations were developed collaboratively by the research team. Findings were then interpreted in relation to the study's sociotechnical, linguistic, practice-oriented, and evaluative analytical perspectives.

The multilingual research team included scholars with Chinese, Japanese, and Mongolian linguistic, cultural, and educational experience. Comparative interpretations were discussed collaboratively, and ambiguous translated passages were checked against the original-language material where necessary. This process was intended to reduce the risk of treating contextual differences as fixed national or cultural characteristics.

A large language model was used during manuscript preparation to support English-language editing, restructuring, and drafting. All AI-assisted text was critically reviewed and revised by the authors, who take responsibility for the final manuscript. Generative AI was not used to analyze interview data or generate empirical findings.

## 5. Findings

### 5.1 Access Pathways, Tool Choice, and Interaction Language

The interviews revealed distinct patterns in how students across the three study contexts accessed GenAI, selected among available tools, and chose languages for interaction. These differences concerned not only which systems students used, but also the technological and linguistic pathways through which GenAI became practically accessible in academic work.

Participants in China described the most diverse platform environment, combining domestic GenAI systems with global platforms such as ChatGPT. Access conditions were closely connected to tool choice. Some participants used virtual private networks (VPNs) to reach global systems, while others relied primarily on domestic tools that were more readily available and adequate for their purposes. Several students moved between platforms according to perceived strengths and task requirements. Chinese was commonly used with domestic systems, while English appeared more often in interactions with some global platforms.

Japanese participants predominantly used globally available systems, especially ChatGPT, and did not report the same access-related constraints described by some Chinese participants. Interaction occurred mainly in Japanese, allowing students to incorporate globally developed GenAI tools into academic activities without routinely switching to English. Although GenAI was also used for English-related tasks, Japanese remained the principal language of everyday interaction.

Mongolian participants likewise relied heavily on globally available systems, but their accounts showed more frequent multilingual interaction. Several explicitly linked language choice to perceived differences in response quality. One participant explained that "the answers in Mongolian were not very good, so I mainly use English now." Others similarly described obtaining more complete or reliable responses in English or Korean than in Mongolian. These accounts suggest that access to a global GenAI platform did not necessarily translate into equally usable interaction through students' primary language.

## 5.2 GenAI across Academic Practices

### 5.2.1 Patterns of GenAI Use

The co-occurrence analysis identified nine recurring patterns of GenAI use: problem solving and information seeking; idea generation and academic paper search or reading; foreign-language translation and correction; learning through comparing and checking outputs; assignment and report writing; image generation and analysis; simplifying difficult content; information collection and summarization; and proofreading and text revision (Figure 1; Table 2). These thematic groupings were developed by interpreting lexical associations in relation to the interview passages rather than treating network clusters as automatically generated themes.

Qualitative comparison further indicated that these practices could be organized into three overlapping forms of engagement: exploration, production, and refinement. Exploratory uses involved approaching unfamiliar topics, locating information, clarifying concepts, and developing ideas. Production involved generating or developing written, translated, or visual materials, while refinement involved checking, comparing, proofreading, and revising existing work. These categories describe different positions that GenAI could occupy within academic activity rather than sequential stages of use.

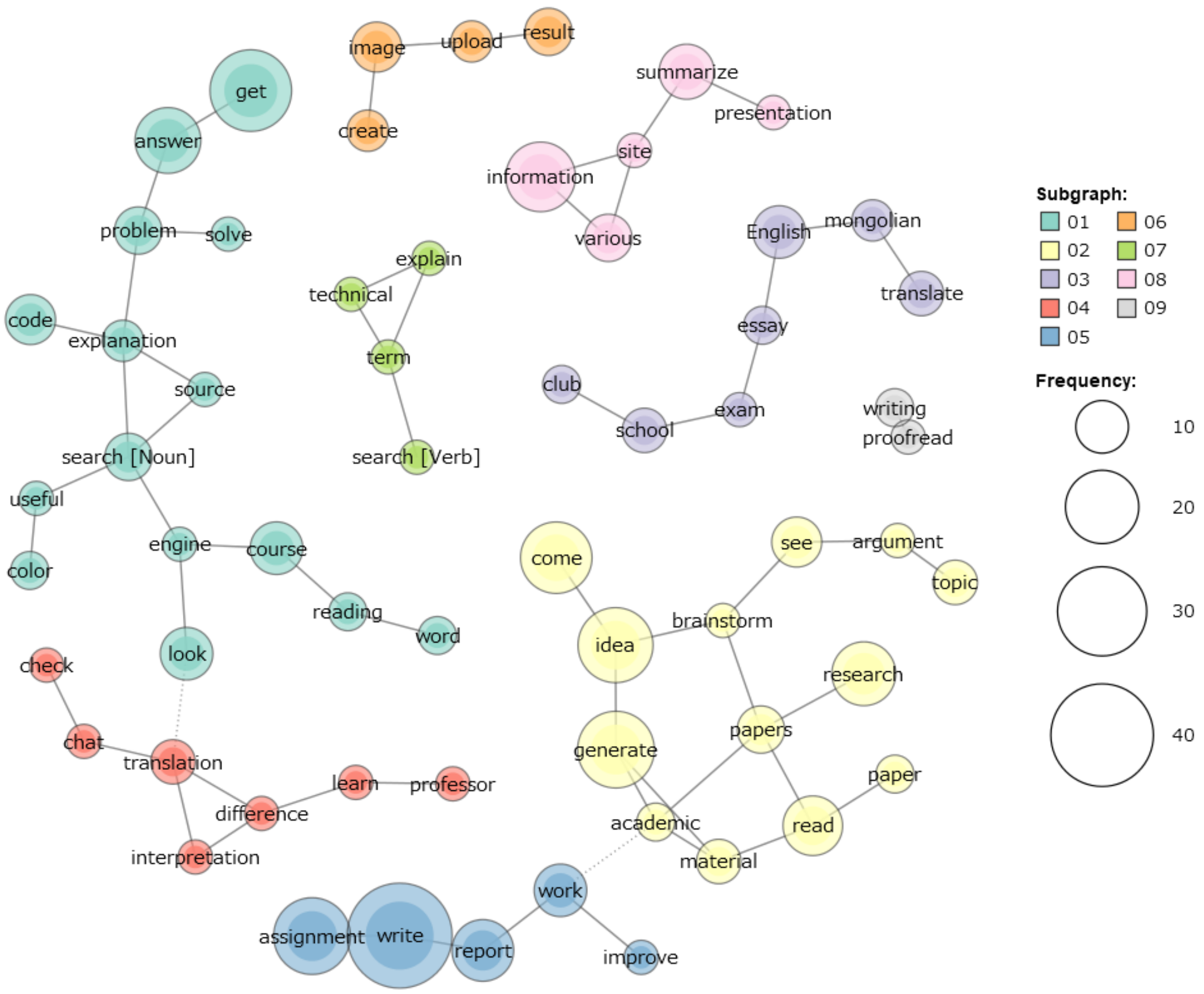

Figure 1. Co-occurrence network of reported GenAI use

Table 2. Thematic interpretation of the GenAI use co-occurrence network

| Theme | Description | Representative extract |
|---|---|---|
| **1. Problem Solving and Information Seeking** | Obtaining answers, explanations, sources, or guidance when encountering academic problems. | J1: "There was an assignment I didn't understand… I used it as one way of solving the problem." |
| **2. Idea Generation and Academic Paper Search/Reading** | Brainstorming ideas, developing research directions, and supporting the search or initial reading of academic materials. | C7: "I wanted to discuss some research ideas with it… it actually provided suggestions that had some reference value." |
| **3. Foreign-Language Translation and Correction** | Translating texts and improving expression in an additional language. | M2: "I first translate from Mongolian to English… then paste that into ChatGPT and ask, 'Will native speakers understand this?'" |
| **4. Learning through Comparing and Checking Outputs** | Comparing student-generated translations or interpretations with AI outputs to support learning. | J4: "I don't copy it directly, but I might notice… maybe I could use this expression instead." |
| **5. Assignment and Report Writing** | Organizing, formulating, or improving written academic work. | C5: "I write my research proposal or personal statement myself first, and then ask it to polish the writing." |
| **6. Image Generation and Analysis** | Generating visual materials or uploading images for interpretation and design support. | M9: "I also use its image-generation function… for things such as generating backgrounds for logo designs." |
| **7. Simplifying Technical Terms and Difficult Content** | Explaining difficult terminology, theories, or course material in more accessible language. | M7: "In philosophy… ChatGPT explained [the topics] using examples from everyday life." |
| **8. Information Collection and Summarization** | Condensing lengthy materials or extracting information relevant to an academic task. | C4: "I now tend to put it directly into AI and ask it to extract the parts related to my research." |
| **9. Proofreading and Text Revision** | Checking and revising student-authored texts after drafting. | J9: "I use it after writing the text myself to check whether I have written anything unnatural." |

*Note. Interview extracts originally provided in Chinese, Japanese, or Mongolian were translated into English. Participant identifiers indicate country: C = China, J = Japan, and M = Mongolia.*

### 5.2.2 Cross-Context Variation in Academic Use

Correspondence analysis was used to examine differences in the relative salience of use-related vocabulary across the three study samples (Figure 2). With three study-context categories, the solution comprised two dimensions. Dimension 1 accounted for 51.4% of the total inertia and Dimension 2 for 48.6%.

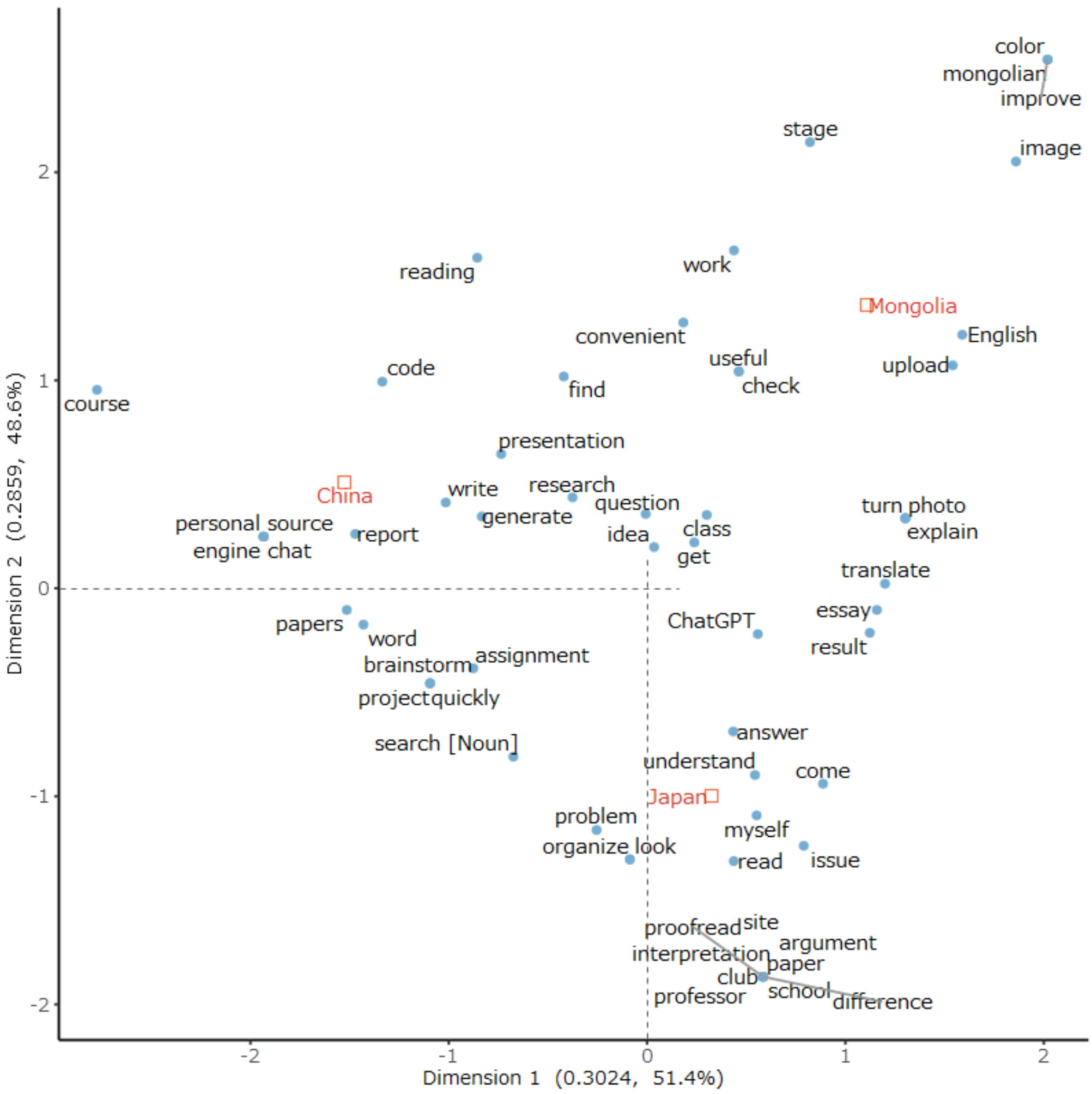

Figure 2. Correspondence analysis of GenAI use by country sample

The Chinese sample was positioned relatively closer to terms such as *presentation, report, source,* and *papers*, indicating greater salience of text-based academic production and information-related activities. The Japanese sample was closer to terms including *interpretation, professor, problem,* and *read*, suggesting greater prominence of understanding, reading, and problem-oriented uses. The Mongolian sample was positioned closer to *English, Mongolian, image,* and *upload*, reflecting the relative prominence of multilingual and image-related practices.

These associations indicate differences in emphasis rather than mutually exclusive forms of use. Many GenAI practices appeared across all three samples, while the correspondence analysis highlighted which forms of vocabulary were relatively more salient in each context. The patterns should therefore be interpreted as contextual tendencies within the present dataset rather than as general characteristics of Chinese, Japanese, or Mongolian university students.

### 5.3 Evaluating GenAI Use: Concerns, Risks, and Boundaries

#### 5.3.1 Patterns of Concern and Evaluation

The co-occurrence analysis identified seven recurring themes in participants' concerns regarding GenAI: privacy and data-use risks; sources, ownership, and cultural fit; inaccurate knowledge and learning dependence; sensitive questions and exposure; harmful outputs and jailbreak prompts; AI-generated images and social values; and decline in critical thinking (Figure 3; Table 3). These themes were developed by interpreting lexical associations in relation to the interview passages rather than treating network clusters as automatically generated categories.

For interpretive synthesis, the seven themes clustered around three broader domains: information and data risks, boundaries of human–AI interaction, and educational and social consequences. Information and data risks included concerns about accuracy, provenance, privacy, and the storage or reuse of information. Boundaries of interaction concerned ownership, culturally inappropriate or incomplete outputs, sensitive information, and the limits of acceptable AI assistance. Educational and social consequences included dependence, reduced engagement with original materials, weakening of independent or critical thinking, and wider cultural or social effects.

Importantly, concerns did not necessarily result in non-use. Participants often continued to use GenAI while adjusting what information they disclosed, checking outputs against other sources, or reconsidering how much academic work should be delegated to AI. Evaluation therefore formed part of students' ongoing engagement with GenAI rather than occurring only after use.

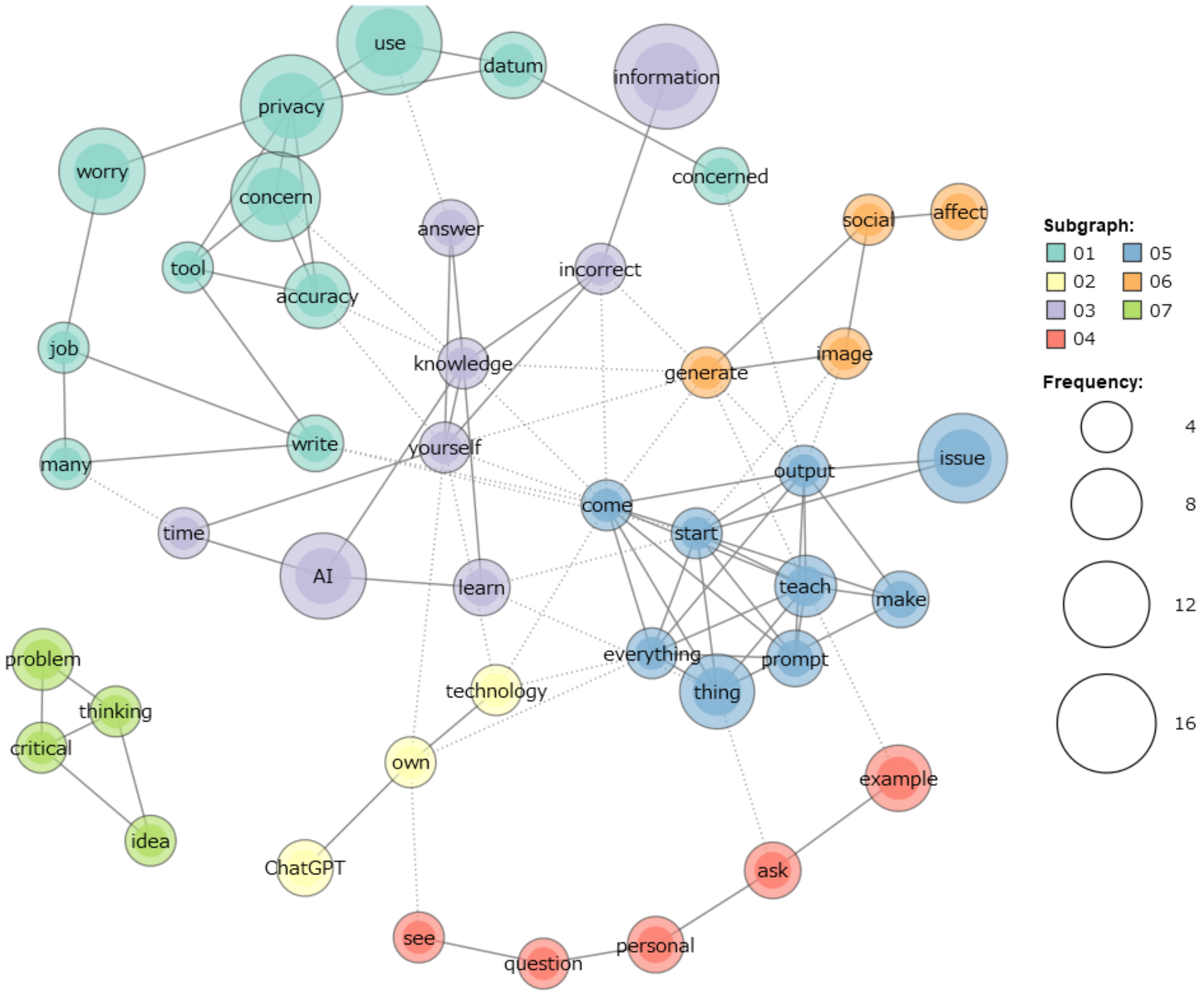

Figure 3. Co-occurrence network of concerns regarding GenAI use

Table 3. Thematic interpretation of the GenAI concern co-occurrence network

| Theme | Description | Representative extract |
|---|---|---|
| **1. Privacy & data use risks** | Collection, storage, reuse, or leakage of information entered into GenAI systems. | M8: "I am still a bit worried about how the data are stored." |
| **2. Sources, ownership & cultural fit** | Unclear sources or ownership and inadequate representation of local linguistic or cultural contexts. | M10: "Foreign platforms haven't really… integrated much about Mongolia… the answers feel incomplete." |
| **3. Inaccurate knowledge & learning dependence** | Inaccurate or incomplete outputs and reduced engagement with original materials. | C4: "It is much faster, but because I lose a lot of the context, my handling of the literature becomes less rigorous and less systematic." |
| **4. Sensitive questions & exposure** | Risks associated with entering highly personal or sensitive information. | M5: "I don't really ask personal things, so I don't worry much about privacy or data leaks." |
| **5. Harmful outputs & jailbreak prompts** | Possibility that safeguards can be circumvented to generate inappropriate or harmful content. | J6: "There is also the issue of jailbreak prompts… prompts that tell it to forget those restrictions." |
| **6. AI images & social values** | Potential distortion of cultural meanings or reproduction of problematic social representations. | C14: AI-generated interpretations could "distort what the culture originally meant and then spread that distorted understanding." |
| **7. Critical thinking decline** | Potential weakening of independent reasoning, creativity, and problem solving through excessive reliance. | J1: "It can reduce people's ability to think—it can take away our thinking ability." |

*Note. The themes represent interpretive categories derived from the co-occurrence patterns and the contexts in which the words appeared. Representative excerpts illustrate each theme and do not imply that the concern was unique to the country shown.*

### 5.3.2 Cross-Context Variation in Concerns

Correspondence analysis was used descriptively to examine the relative salience of concern-related vocabulary across the three study samples (Figure 4). With three study-context categories, the solution comprised two dimensions. Dimension 1 accounted for 51.62% of the total inertia and Dimension 2 for 48.38%.

The Chinese sample was relatively more closely associated with terms concerning information, accuracy, privacy, and the representation of knowledge, including questions about the adequacy of AI-generated information in local or culturally specific contexts. The Japanese sample showed greater salience of concerns about the boundaries of AI assistance, learning, and critical thinking, particularly the possibility that excessive delegation could reduce independent reasoning. The Mongolian sample was relatively more associated with practical reliability, language adequacy, and local relevance, including concerns about incomplete Mongolian-language responses and the need to verify outputs or switch languages.

These differences represent variation in emphasis rather than mutually exclusive categories of concern. Accuracy, privacy, dependence, and reliability appeared across all three samples but were articulated in relation to different technological and linguistic circumstances. The correspondence patterns should therefore be interpreted as contextual tendencies within the present dataset rather than as fixed characteristics of Chinese, Japanese, or Mongolian students.

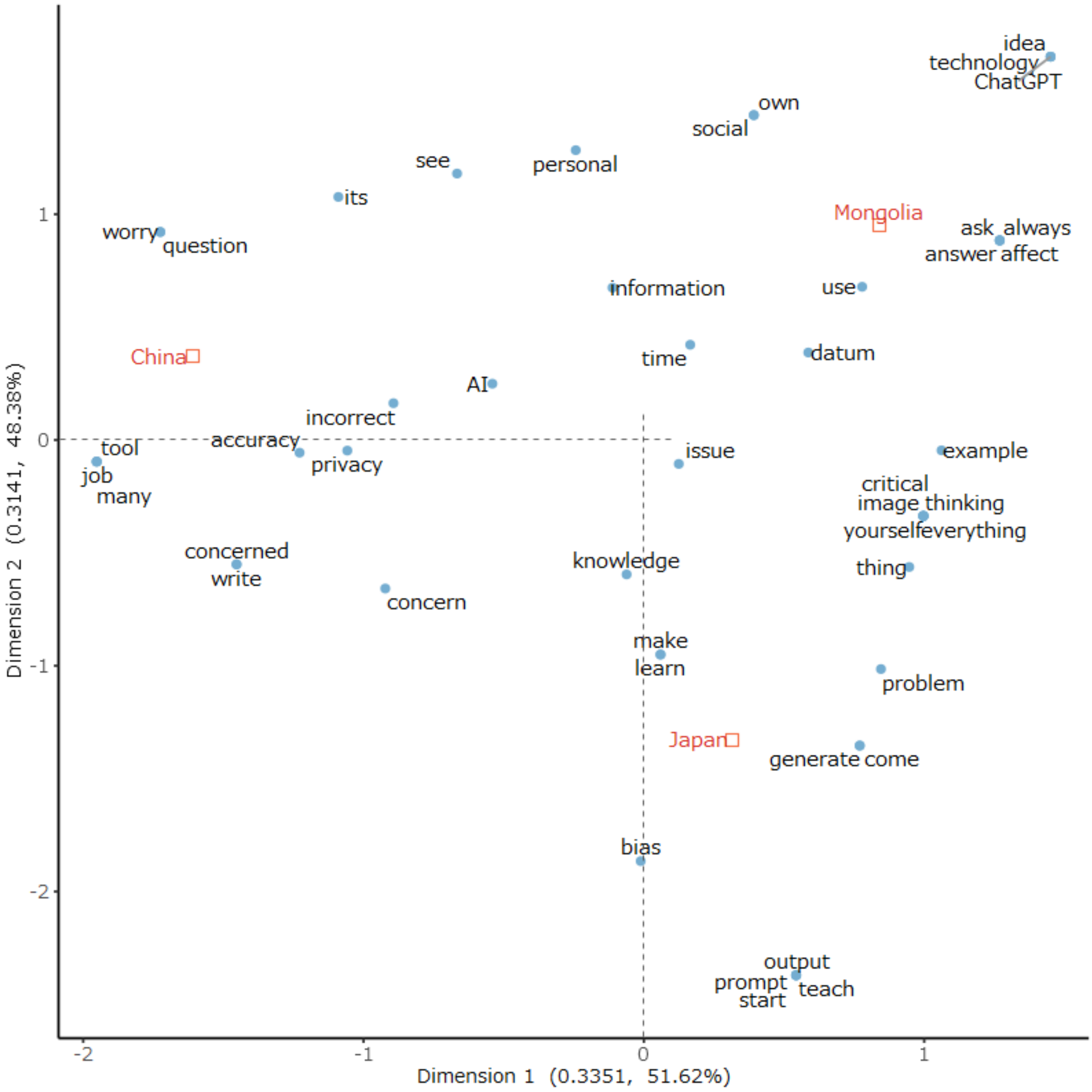

Figure 4. Correspondence analysis of GenAI concerns by country sample

## 6. Discussion

### 6.1 GenAI as an Embedded Academic Practice

Students' accounts show that GenAI was embedded at multiple points in academic work rather than functioning primarily as a tool for generating finished products. Across the interviews, GenAI supported exploration, production, and refinement. These forms of engagement were overlapping rather than sequential, and the same technological function could assume different roles depending on where and how it entered an academic task. Translation, for example, could serve as direct text production or as a comparative learning activity, while writing assistance ranged from generating initial material to polishing student-authored text.

This finding reinforces research showing that student engagement with GenAI is heterogeneous and process-dependent. Prior studies have identified different patterns of reliance, agency, and responsibility in GenAI-supported learning (Stojanov et al., 2024; Yang et al., 2024; Hou et al., 2025; Zhao et al., 2024). The present study extends this line of work by foregrounding the position and function of GenAI within academic activity. Whether AI is used before, during, or after students' own intellectual work may be more educationally informative than the simple presence or frequency of use.

A relational affordance perspective helps explain this distinction. Educational possibilities emerge through the relationship among technological capabilities, learners' purposes and abilities, and the activities in which the technology is used (Antonenko et al., 2017). Summarization, translation, or text generation therefore does not carry a fixed educational meaning. Its significance depends on how the capability is taken up within a particular academic practice and how responsibility is distributed between the student and the system.

The findings therefore suggest that “GenAI use” is too broad a category to capture its educational significance. A more useful question is where GenAI enters academic work, what role it performs, and how responsibility is distributed between the student and the system. This practice-oriented interpretation provides a basis for examining how such forms of engagement are shaped by sociotechnical and linguistic conditions.

### 6.2 GenAI Engagement as Linguistically and Sociotechnically Situated

The comparative findings show that students’ engagement with GenAI was shaped not only by platform availability but also by the linguistic resources through which those platforms could be used. The three study contexts presented different configurations. Chinese participants navigated between domestic and global GenAI systems, with accessibility and perceived platform strengths influencing tool choice. Japanese participants predominantly used globally available systems through Japanese. Mongolian participants also relied heavily on global systems, but several shifted from Mongolian to English or other languages when they perceived non-Mongolian responses as more complete or reliable. These patterns highlight an important distinction between platform access and linguistic usability.

This interpretation connects two strands of emerging research. Multilingual and translanguaging studies emphasize that GenAI interaction is shaped by users' linguistic repertoires and by tensions surrounding linguistic inclusion (Jeon et al., 2025; Ou et al., 2024; Wang, 2024). Computational research likewise shows that LLM performance can vary across languages and educational tasks, with lower-resource languages performing less strongly in some settings (Gupta et al., 2025). The present findings connect these strands from the perspective of student practice by showing how some students responded to perceived linguistic asymmetries through language switching, output comparison, and translation-supported interaction.

Language should therefore be understood as more than either a learner characteristic or a specific application of GenAI. Drawing on linguistic mediation (Lantolf, 2006), the findings suggest that language forms part of the interactional pathway through which students formulate prompts, interpret responses, and determine whether outputs are usable for academic purposes. In the Chinese sample, linguistic resources operated alongside access to domestic and global platforms, consistent with prior research on accessibility, language proficiency, disciplinary needs, and perceived platform capabilities (Xie et al., 2025). In the Mongolian sample, language switching illustrates how additional linguistic resources may become important when a preferred interaction language does not produce satisfactory results.

These patterns should not be interpreted as inherent national differences. Rather, they reflect the particular combinations of platform environments and linguistic resources encountered by participants in the three study sites. Bringing sociotechnical conditions and linguistic mediation together shifts attention from formal access alone toward the conditions under which GenAI becomes practically usable for academic work.

### 6.3 Evaluating and Negotiating the Boundaries of GenAI Use

Students' concerns did not translate into simple acceptance or rejection of GenAI. Participants continued to use these systems while questioning accuracy, privacy, ownership, dependence, cultural representation, and possible effects on critical thinking. Some responded by checking information against other sources, comparing outputs, limiting personal disclosure, or reconsidering how much of an academic task should be delegated to AI. Evaluation therefore formed part of ongoing GenAI use rather than occurring only after interaction.

A central tension concerned efficiency and intellectual responsibility. Capabilities such as summarization, explanation, translation, and text generation could support academic work, but they also became problematic when they appeared to replace activities considered important for learning. Summarization could reduce engagement with original sources, immediate answers could discourage sustained problem solving, and text generation could shift responsibility for academic production away from the student. The educational significance of GenAI therefore depended partly on how students judged the appropriate extent of AI assistance.

This finding aligns with AI-literacy perspectives that emphasize critical evaluation and ethical awareness rather than technical proficiency alone (Long & Magerko, 2020; Ng et al., 2021). At the same time, the present study suggests that evaluative judgment is not simply an abstract individual competency. It is exercised within specific academic practices and conditions, as students assess reliability, negotiate disclosure, and reconsider reliance in response to particular tasks and outputs.

The emphasis of these judgments varied across the three study samples. Chinese participants more often foregrounded informational and sociocultural implications, Japanese participants the boundaries of AI assistance and independent judgment, and Mongolian participants the relationship among reliability, language adequacy, and local relevance. These differences should be understood as contextual tendencies rather than national characteristics. More importantly, they show that students' evaluations of GenAI were themselves situated within the technological, linguistic, and academic conditions through which use occurred.

### 6.4 From Access to Realized Affordances: An Integrative Framework

Taken together, the findings suggest that technological access alone provides an incomplete account of students' engagement with GenAI. The three study contexts differed not only in platform availability, but also in the linguistic resources through which students interacted with GenAI, the academic practices into which it was incorporated, and the judgments students made about the reliability and appropriateness of AI assistance. Building on these empirical patterns and the analytical perspectives introduced earlier, the study develops an integrative framework in which student engagement with GenAI is understood as **situated, mediated, enacted, and evaluated**. The framework represents an empirical and conceptual synthesis rather than a predetermined coding structure or causal model.

**Situated** refers to the sociotechnical conditions within which students encounter GenAI, including platform availability, access constraints, and the configuration of domestic and global systems. **Mediated** highlights the linguistic resources through which students formulate prompts, interpret outputs, and move across languages when necessary. **Enacted** concerns how technological possibilities are incorporated into academic practices such as exploration, production, comparison, and revision. **Evaluated** refers to students' ongoing judgments about accuracy, privacy, authorship, dependence, cultural adequacy, and the appropriate boundaries of AI assistance. These four aspects are analytically distinguishable but empirically interconnected.

This synthesis clarifies what is meant here by **realized affordances**. GenAI systems may technically offer capabilities such as explanation, translation, summarization, or text generation, but these possibilities do not automatically become educationally usable simply because a platform is accessible. Rather, their realization depends on the interplay among sociotechnical conditions, linguistic resources, academic practices, and evaluative judgments. *Realized affordances* therefore refer to the educational possibilities of GenAI that become practically usable within particular academic contexts. These relationships are summarized in Figure 5.

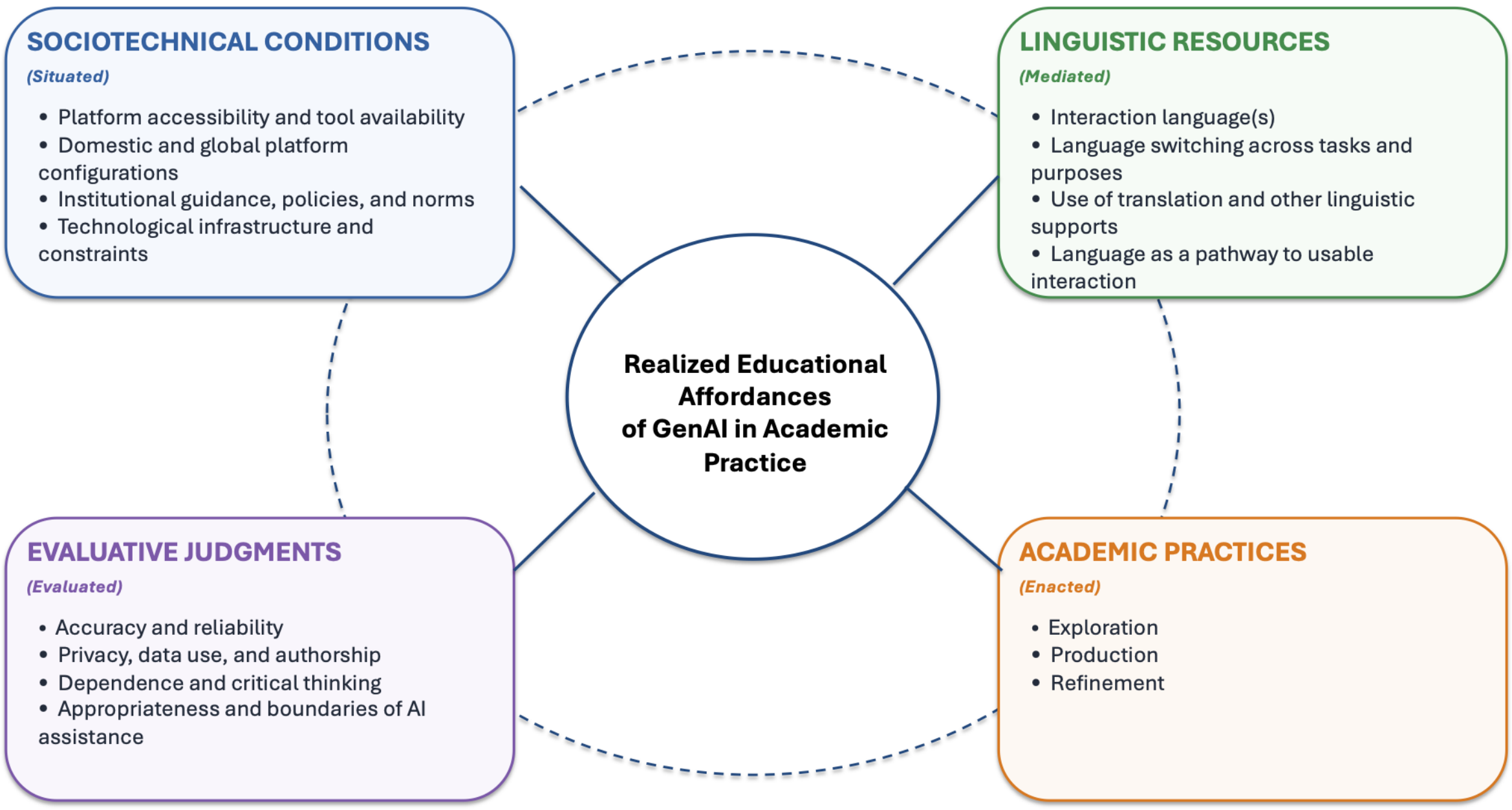

Figure 5. From access to realized affordances: An empirically informed integrative framework of university students' engagement with GenAI.

The framework does not imply a linear sequence from access to mediation, enactment, and evaluation. In practice, these processes are iterative and mutually shaping. An unsatisfactory output may prompt a student to change language, switch platforms, revise a prompt, verify information elsewhere, or reduce reliance on the system. Evaluation can therefore reshape both mediation and enactment, just as platform conditions can constrain the linguistic and academic possibilities available to students. The central proposition is that GenAI affordances are not realized through technological access alone; their educational realization emerges through the situated interplay of sociotechnical conditions, linguistic resources, academic practices, and students' evaluative judgments.

The framework therefore shifts attention from whether students have access to GenAI toward what forms of educational possibility become usable under particular conditions. It also provides a basis for comparing GenAI engagement across multilingual and sociotechnically diverse higher education settings without treating national context itself as the explanation for observed differences.

### 6.5 Implications for Higher Education

The findings have several implications for how higher education institutions approach GenAI integration. First, institutional discussions of access should extend beyond whether students can technically reach a particular platform. As the comparison illustrates, practical access is also shaped by the linguistic resources students can draw on and by the quality of interaction available through different languages. Institutions working in multilingual settings should therefore consider linguistic usability, local-language support, and students' need to move across languages or tools when evaluating the accessibility of GenAI-supported learning.

Second, GenAI literacy should extend beyond technical proficiency or prompt-writing skills. Students need opportunities to develop evaluative judgment concerning the reliability and provenance of AI-generated information, privacy and disclosure, authorship, linguistic variation in outputs, and the appropriate delegation of academic work. Such judgment is particularly important because the educational value and risks of GenAI often arise from the same capabilities: summarization can support efficient exploration but also reduce engagement with original sources, while text generation can facilitate production but also shift intellectual responsibility away from the learner.

Finally, institutional guidance may be more useful when it is connected to particular academic practices rather than framed only as general permission or prohibition. Guidance for exploratory use, writing support, translation, revision, and assessment may require different expectations regarding verification, disclosure, authorship, and student responsibility. A context-sensitive approach can therefore help institutions support meaningful GenAI use while recognizing that the educational possibilities students are able to realize may differ across technological and linguistic environments.

## 7. Conclusion

This study examined how university students engage with GenAI across three contrasting linguistic and sociotechnical higher education contexts in China, Japan, and Mongolia. The findings show that GenAI was embedded across exploration, production, and refinement of academic work rather than confined to content generation. At the same time, students encountered different technological and linguistic pathways to GenAI: participants in China navigated between domestic and global systems, those in Japan predominantly used globally available systems through Japanese, and those in Mongolia more frequently shifted toward English or other languages when Mongolian-language outputs were perceived as less satisfactory. Students also continued to use GenAI while evaluating its reliability, privacy implications, effects on intellectual independence, and appropriate boundaries of assistance.

Building on these findings, the study develops an empirically informed integrative framework in which student GenAI engagement is understood as situated, mediated, enacted, and evaluated. The framework highlights that technological access alone does not determine the educational possibilities students are able to use. Rather, realized affordances emerge through the interplay of sociotechnical conditions, linguistic resources, academic practices, and students' evaluative judgments. This perspective draws particular attention to linguistic mediation: access to the same or similar GenAI platforms may not translate into equivalent educational usability when students encounter different levels of support or output quality across languages.

These findings suggest that higher education institutions should approach GenAI integration as more than a question of platform provision or technical proficiency. Supporting meaningful use also requires attention to linguistic accessibility, task-specific academic practices, and students' capacity to judge when AI assistance supports rather than substitutes for intellectual engagement.

The study has several limitations. It draws on 42 undergraduates from one university in each study context and therefore does not represent national student populations. The computer-assisted analyses were also conducted on translated interview corpora and are influenced by corpus composition, translation, and preprocessing decisions. In addition, data were collected at different times and through somewhat different interview modes across the three sites. Future research should examine the transferability of the proposed framework with larger and more diverse samples, additional linguistic contexts, and longitudinal or mixed-method designs capable of tracing how GenAI practices evolve as technologies and local-language support continue to change.